\documentclass[preprint,aps,showpacs]{revtex4}
\def\ee{\end{equation}}
\def\be{\begin{equation}}
\def\ba{\begin{eqnarray}}
\def\ea{\end{eqnarray}}
\def\ll{\label}

\begin{document}

\title{Statistical mechanics of thermal contact between \\
system and bath with long-range interactions}
\author{Ramandeep S. Johal$^1$} \email{johalr69@yahoo.co.in}
\author{Renuka Rai$^{2}$} \email{rrai@lycos.com}
 \affiliation{$^1$Post-Graduate Department of Physics,
  Lyallpur Khalsa College, Jalandhar-144001, India. \\
$^{2}$Post-Graduate Department of Chemistry, Lyallpur Khalsa College,
 Jalandhar-144001, India.}
%


\begin{abstract}
In this paper, we address the possibility of generalising the standard analysis 
of thermal contact between a sample system and a
heat bath, by including long range interactions between them.
As a concrete example, both system and bath are treated within the 
long range Ising model. For this model, we derive the
equilibrium probability distribution of the energy of the sample system. 
Equilibrium properties of the system magnetisation and
stability of the solutions is discussed. We find existence of a metastable
phase below a critical temperature of the bath.
\end{abstract}

\maketitle

\section*{I. INTRODUCTION}
The theory of equilibrium thermodynamics and statistical mechanics  has
 been  rigorously developed only for the case of short range interactions
among the components of the system. The presence of long range interactions 
among different components, makes a system much more complex and the standard 
approaches become inapplicable. Currently, there is a lot of 
interest to develop new methods and tools to deal with systems involving
long range interactions \cite{Dauxois2002}.
Apart from the standard examples within the fields of
cosmology and astrophysics \cite{Padmanabhan1990}, 
 a growing number of physical laboratory systems have recently 
emerged in which the interactions are long-ranged, notably in the areas
of plasma physics \cite{Elskens2000}, nuclear physics and atomic clusters
\cite{Chomaz2004}, Bose-Einstein condensates \cite{Arimondo2002},  
2D hydrodynamics \cite{Laval2001, Chavanis2005}, magnetism 
\cite{Ispolatov2000,Barre2001,Ellis2004,Costeniuc2005} 
and so on (see \cite{Dauxois2002} and references therein). 
In systems with long range interactions, energy is generally nonadditive. 
Due to this feature, the thermodynamics of 
such systems displays unusual properties, like inequivalence of different 
ensembles \cite{Eyink1993, Ellis2000}, 
possible temperature discontinuity at first order
transitions and negative specific heat in microcanonical ensemble 
\cite{Lynden1999}.

Now the standard derivation of Boltzmann-Gibbs distribution considers a thermal
contact between a sample system and a heat bath
with a very large heat capacity.
The derivation assumes an additive property 
for energy,  to arrive at the desired result. The case of long range
interactions between the system and bath seems to be beyond the scope of this
derivation \cite{Dauxois2002}. For one, to include such interactions, 
we have to specify the nature of the interactions between the components
of the system and the bath. This implies specifying the microscopic structure
of the bath also which makes the characterisation of the bath more detailed.
In contrast, within canonical ensemble the bath is characterised only by
its temperature irrespective of its microscopic model. Further, nonadditivity
of energy brings additional difficulties.  
 
In this paper, we consider this latter case and derive the probability 
distribution for a system in thermal contact with a heat bath in the
presence of long range interactions. In section II, 
we take  a specific
model for long range interactions of  mean-field type  between the system and
the bath, and present such a derivation.  In section III, we compare our model
with the Curie-Weiss model which is paradigmatic model for ferromagnetism.
In section IV, equilibrium value of the magnetisation is discussed and 
stability of the solutions is highlighted. The last section V, presents
a summary of our approach and results.

\section*{IIA. Canonical Ensemble from Microcanonical Ensemble}
For the sake of completeness and to fix the notation, let us review one of the
text-book derivations of the canonical ensemble \cite{Pathria1996}.
Denote by $E_1$ the energy  of the sample and by $E_2$ the energy
of the reservoir.  The sample and the  reservoir together form
an isolated system and the interactions between them are usually considered to
be short ranged. Thereby the total energy of the composite system say $E_0$,
is given by  the sum of energies $E_1+E_2$ and is  constant.
Next, $\Omega_{i}(E_{i})$ where $i=1,2$, denote
the    respective number   of   microstates at the given values of energies. The
probability that the system 1 is in a certain microstate of energy $E_1$ is
given by

\begin{equation}
p_1(E_1)=\frac{\Omega_1(E_1)\; \Omega_2(E_2)}{\Omega_{1+2}(E_0)},
\label{firsteq}
\end{equation}

\noindent
where $E_2 = E_0 - E_1$, and $\Omega_{1+2}(E_0)$  is the total  number
of states  available for the composite system $1+2$. Thermodynamic entropy of
 the reservoir is given by Boltzmann's formula:

\begin{equation}
S_2(E_2) = k_{\rm B} \ln \Omega_2(E_2).
\ll{s2log}
\end{equation}

\noindent
Therefore, $p_1(E_1) \propto {\Omega_1(E_1)\; e^{S_2 / k_{\rm B}  }}$. In the
following, we keep by $k_{\rm B} =1$ for simplicity.
Now expand $S_2(E_2)$ around the most probable value $\bar{E}_2$ as:

\begin{equation}
S_2(E_2)  = S_2(\bar{E}_2) +  \left( \frac{d S_2}{d E_2} \right)_{\bar{E}_2}
 (E_2 - \bar{E}_2) + \cdots .
\label{expands2}
\end{equation}

\noindent
As the assumed size of the bath is quite large, its instantaneous energy
will be very close to the most probable energy. Therefore,
taking $E_2 - \bar{E}_2$ to be small, the higher order terms
in the above expansion are negelected. Let the inverse temperature parameter
$\left( {d S_2}/{d E_2} \right)_{\bar{E}_2} = {\bar{\beta}}_2$.
Then using additive property of the energy,
we have  $E_2 - \bar{E}_2 = \bar{E}_1 - E_1$ and thus we obtain the Boltzmann
distribution for system energies: $p(E_1) \propto  \Omega_1(E_1)\;
\exp(-{\bar{\beta}}_2 E_1)$.

\section*{IIB. ISING MODEL WITH LONG RANGE INTERACTIONS}
Now consider the Ising spin model for both the system and the bath 
interacting with each other through long
range interactions. (For a recent discussion of mutual thermal equilibrium
and alternate thermodynamic descriptions within this model, see Ref. 
\cite{Johal2006}.)

Take a lattice of $N$ spins with the hamiltonian  given by

\be
E = - \sum_{j} h_j \; \sigma_j - \sum_{< j k >} {\cal J}_{jk} \; \sigma_j
 \; \sigma_k.
\ll{hamilg}
\ee

\noindent
Each spin variable $\sigma_i$ is defined over the set $\{+1, -1\}$.
The total system is assumed to be separated into two regions representing
a sample system and a heat bath.
The spin excess in each region is given by $e_i = N_{i}^{+}  - N_{i}^{-}$,
where $N_{i}^{\pm}$ are the number of up (+) and down (-) respectively.
${\cal J}_{j k}$ and $h_j$ are the known coupling constants within each region
which for simplicity, are taken to be constant over a region.
We focus on the ferromagnetic interactions which mean ${\cal J}$'s are
greater than zero. In terms of new variables $E_i = - h_i e_i$, sometimes called
the {\it local energies}, the total energy upto a constant term,
can be written in the form

\be
E =   E_1 + E_2 - a {E_1}^2 - b {E_2}^2 - c E_1 E_2,
\ll{hamilse}
\ee

\noindent
where $a, b$ and  $c$ are suitably defined constants to be elaborated later.
Clearly, in the presence of long ranged interactions, the
energies as well as the temperatures of the systems, get modified. 
For example, the most probable energy 
of system 1 is ${\cal E}_1 = \bar{E_1}- a \bar{E_1}^2$ and its inverse
temperature is

\be
\beta^{*}_{1} \equiv
\frac{d S_1}{d {\cal E}_1} =  \frac{{\bar{\beta}}_1}{ (1 - 2a\bar{E_1})},
\label{bet1*}
\ee

\noindent
where $\bar{\beta}_1 = d S_1 /{d \bar{E}_1}$ is the inverse temperature in the
absence of long range interactions within system 1. Similarly, for the bath 
we have

\be
\beta^{*}_{2} =  \frac{{\bar{\beta}}_1}{ (1 - 2b\bar{E_2})}.
\label{bet2*}
\ee

In the following, we adapt the scheme as given in Eqs. 
(\ref{firsteq})-(\ref{expands2}) to the case when the total energy is given by
(\ref{hamilse}). It is convenient to express the entropy of a system
in terms of the respective local energy $E_i$. To evaluate $(E_2 - \bar{E}_2)$ 
for this case, we proceed as follows:

In terms of the most probable values of the local energies, we have

\be
E =   \bar{E}_1 + \bar{E}_2 - a \bar{E_1}^2 - b \bar{E_2}^2
      - c \bar{E}_1 \bar{E}_2.
\ll{hamilse2}
\ee

\noindent
Equating Eqs. (\ref{hamilse}) and (\ref{hamilse2}) and defining
$\Delta E_i = E_i - \bar{E}_i$, $i=1,2$, we can write

\be
\Delta E_1 + \Delta E_2 - a( {E_1}^2 - \bar{E_1}^2 )
-b ( E_2 + \bar{E}_2 ).\Delta E_2 - c (E_1 E_2 - \bar{E}_1 \bar{E}_2) = 0.
\label{sube}
\ee

\noindent
Now due to the large size of the system 2, the energy $E_2$ is very closely
approximated by its most probable value. Thus the following terms in the
above equation can be approximated as

\be
(E_2 + \bar{E}_2).\Delta E_2 \simeq 2 \bar{E}_2 . \Delta E_2,
\ee

\noindent
and

\be
(E_1 E_2 - \bar{E}_1 \bar{E}_2) \simeq \bar{E}_2 . \Delta E_1.
\ee
Using the above relations in Eq. (\ref{sube}), we get the expression

\be
\Delta E_2 = \frac{- \Delta E_1 (1-c \bar{E}_2) + a ({E_1}^2 - \bar{E_1}^2)
}{(1-2b \bar{E}_2)}.
\label{de2}
\ee

\noindent
Finally, substituting (\ref{de2}) in (\ref{expands2}), we arrive at the
probability distribution for energies of system 1,

\be
p(E_1) = \frac{\Omega_1(E_1) \; e^{ - \beta^{*}_{2}
 \{  (1- c\bar{E}_2) E_1 - a{E_1}^2 \} }}{\cal Z},
\label{pe1}
\ee

\noindent
where

\be
{\cal Z} = \sum_{ E_1} \Omega_1(E_1) \exp \left(- \beta^{*}_{2}
 \{  (1- c\bar{E}_2) E_1 - a{E_1}^2 \} \right),
\label{partitionf}
\ee

\noindent
the sum being over the range of energies accessible to system 1.
It is appropriate here to explain the various paramters above:

\be
a = \frac{J_1}{2 N_1 {h_1}^2}, \quad b = \frac{J_2}{2 N_2 {h_2}^2},
\quad c = \frac{J_{12}}{N_2 {h_1}{h_2}}.
\label{abc}
\ee
The scaling of parameters by the number of particles is in the
same spirit as Kac's priscription \cite{Kac1963} and it guarantees
that the effective hamiltonian (see Eq. (\ref{hpy1}) below) is extensive and 
thus a proper thermodynamic limit is ensured.
\section*{III. ANALOGY WITH CURIE-WEISS MODEL}
Clearly, when there are no interactions with or within the bath, i.e.
$J_2 = J_{12} =0$ implying $b = c= 0$, then we obtain the probability
distribution of the Curie-Weiss model \cite{Honerkamp2002} 
which is described within canonical ensemble with
hamiltonain $H = E_1 - a {E_1}^2$ and the bath at inverse temperature
$\bar{\beta}_2$.

As mentioned above, the presence of long range interactions within the bath has
the effect of modifying the bath temperature to $\beta^{*}_{2}$. On the other
hand, the long range interactions between the system and
the bath are incorporated effectively as a
modified external magnetic field. This can be clearly seen by noting
that the distribution (\ref{pe1}) corresponds to an effective hamiltonian
$H^{\prime} = {(1- c \bar{E}_2)} E_1 - a {E_1}^2$. 
In terms of the magnetisation per spin, $y_i =
\sum_i \sigma_i /N_i = -E_i /(h_i N_i)$ and using definitions (\ref{abc})
for the parameters, we can write

\be
H^{\prime}(y_1) = -N_1 \left[{(h_1 + J_{12} \bar{y}_2)} y_1 + \frac{1}{2} J_1
{y_1}^2 \right],
\label{hpy1}
\ee

\noindent
indicating the extensive property of the effective hamiltonian.
Note that the effective applied field is
$(h_1 + J_{12} \bar{y}_2) \equiv (1- c \bar{E}_2)h_1$, where $h_1$
is the actual applied field. We note that in contrast to the canonical
ensemble, in the present case, the bath has a specific microscopic realisation.
The expression for
the entropy of bath given by Eq. (\ref{s2log}), corresponding to the
 most probable value of magnetisation per spin $\bar{y}_2$ is

\be
S_2( \bar{y}_2 ) = -N_2 \left[ \frac{(1+\bar{y}_2)}{2} \ln 
\frac{(1+\bar{y}_2)}{2}
+ \frac{(1 - \bar{y}_2)}{2} \ln \frac{(1 - \bar{y}_2)}{2} \right].
\label{ents2}
\ee

The most probable energy of the bath given by  ${\cal E}_2 = \bar{E_2}
- b \bar{E_2}^2$, can  also be written as ${\cal E}_2(\bar{y}_2)
= -N_2 (h_2 \bar{y}_2
+ \frac{1}{2} J_2 {\bar{y}_2}^2)$. Then the inverse temperature of bath is

\ba
 \beta^{*}_{2} & = & \frac{dS_2}{d\bar{y}_2}
                     \left(\frac{d{\cal E}_2}{d\bar{y}_2}\right)^{-1}
                     \nonumber \\
                & = & \frac{1}{2(h_2 + J_2 \bar{y}_2)}
                       \ln\left( \frac{1+ \bar{y}_2}{1-\bar{y}_2}\right).
\label{batht}
\ea

\noindent

Note the limiting values for the bath temperature:

(i) For $\bar{y}_2 \to 1$, we have $\beta^{*}_{2} \to \infty$, which corresponds
 to bath temperature of zero degree Kelvin.

(ii) For   $\bar{y}_2 \to 0$, $\beta^{*}_{2} \to 0$,
or in other words, the bath is most disordered at arbitrarily
high temperatures.

\section*{IV. EQUILIBRIUM PROPERTIES AND STABILITY}

It is straightforward to calculate the
free energy $A(y_1)$ from the partition function by the standard methods of
Hubbard-Stratonovich transformation or the saddle point approximation.
Thus we obtain

\be
A(y_1) =   \frac{1}{2} J_1 {y_1}^2 - \frac{1}{\beta^{*}_{2}} \ln \left[
2 \cosh \{ \beta^{*}_{2} (h_1 + J_{12} \bar{y}_2 + J_{1}{y}_1) \}
\right].
\label{freeenergy}
\ee

\noindent
The stationarity condition $\partial A/ \partial y_1=0$ yields a
self-consistent equation
for the equilibrium magnetisation per spin of system 1:

\be
\bar{y}_1 = \tanh \left[ \beta^{*}_{2} (h_1 + J_{12} \bar{y}_2 +
 J_{1}\bar{y}_1) \right].
 \label{meanfeq}
\ee

\noindent
Usually, the analogue of this equation in the Curie-Weiss
model is analysed for $h_1=0$ case to infer the existence of
a critical temperature, above which the system is paramagnetic and
below which it is ferromagnetic. In the ferromagnetic
phase, two values of magnetisation $\pm |\bar{y}_1|$ are
equally allowed, which actually reflects the symmetry of the
hamiltonian.

In the present case, even in the absence of external field ($h_1=0$), there
is an effective magnetic field $J_{12}\bar{y}_2$ due to long range
interaction between system and the bath. Thus at high temperatures,
we have a unique minimum of free energy at a non-zero value of magnetisation.
Thus the sample is magnetised even at high temperatures. As the temperature
is lowered, a metastable solution for magnetisation also appears alongwith the
global equilibrium solution. The limit of metastability (defined by the inverse 
temperature
$\beta^{({\rm m)}}_{2}$ at which the metastable state appears while lowering
 the temperature of the bath) can be calculated by finding the
point of inflexion, where both first and second order derviatives of free energy
vanish. The latter condition yields
\be
\cosh \left[ \beta^{({\rm m)}}_{2} (h_1 + J_{12} \bar{y}_2 +
 J_{1}\bar{y}_1) \right] = \sqrt{ \beta^{({\rm m)}}_{2} {J_1} }.
 \label{eqcosh}
\ee
Using the above condition alongwith Eq. (\ref{meanfeq}), we can show that
\be
\bar{y}_1 = \pm \sqrt{1-\frac{1}{\beta^{({\rm m)}}_{2} J_1}}
\label{y1meta}
\ee
From this equation, it is clear that ${\beta^{({\rm m)}}_{2} J_1} \ge 1$.
The condition, ${\beta^{({\rm m)}}_{2} J_1}= 1$ implies that $\bar{y}_1=0$.
But for $h_1=0$, this stationary solution is satisfied only for $J_{12}=0$,
(see Eq. (\ref{meanfeq}) above). On the other hand, for
$J_{12} \ne 0$, the stationary solution is non-zero.
Thus we see that the condition (\ref{y1meta})
can be satisfied by $\bar{y}_1 \ne 0$ for an inverse temperture
$\beta^{({\rm m)}}_{2}$ which implies that the bath temperature
is lower than the critical temperature for the usual para-Ferro transition
which obeys ${\beta^{({\rm crit)}}_{2} J_1}= 1$. The actual
temperature for limit of metastability may be calculated from
Eqs. (\ref{meanfeq}) and (\ref{y1meta}).

\section*{V. SUMMARY}
We have introduced a new kind of ensemble to generalise the usual
treatment of a thermal contact between
system and heat bath by including long range interactions between them.
The derivation is motivated by the standard 
derivation of the canonical ensemble from the microcanonical
ensemble. However, the crucial difference is the lack of
additivity of the energy. This does not yield exponential
Boltzmann distributions. The form of the distributions is
strongly dependent on the form of hamiltonian of the total
sample plus bath system. Thus incorporating long range
interactions makes the problem more involved as we have to
specify a microscopic model for the bath as well as the
nature of long range interactions between the system and the bath.
In this paper, we have treated a very simple case
when the interactions can be modelled by long range 
Ising model. An appropriate scaling of the interaction
parameters with system or bath size helps to obtain a thermodynamic
limit for the system properties. The observations for this
analysis are that the usual para-ferro transition appears
to be suppressed, with the presence of a net magnetisation
for the system at high temperatures, even in the absence
of an applied field. This happens because the long range
interaction with the bath provides an effective magnetic field.
Moreover, as the temperature is lowerd, there appears a
metastable state. This happens at a temperature which
is generally  lower than the critical temperature of 
para-ferro transition in the Curie-Weiss model. It is hoped that the  approach 
presented in this paper will motivate further studies with other model
long-range interactions, such as slowly decaying interactions. This may 
facilitate the characterisation of 
thermodynamic behaviour in systems when interactions
with the bath cannot be neglected.

\section*{Acknowledgements}
RSJ gratefully acknowledges useful comments on the initial draft of this
paper from Prof. Stefano Ruffo.


\begin{figure}
\input{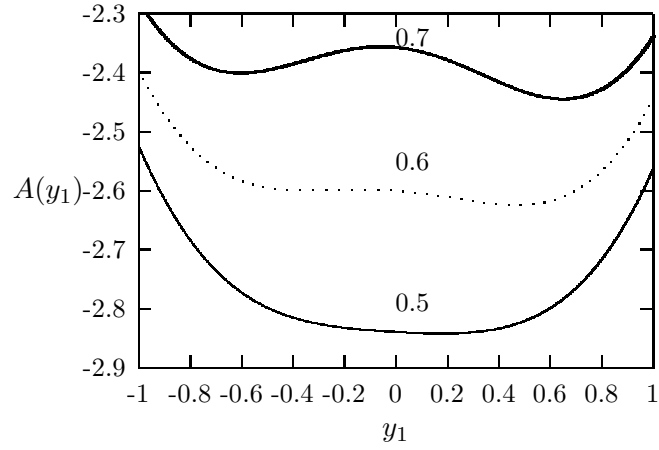}

\caption{Fig. 1: Free energy (Eq. (\ref{freeenergy})) plotted 
against system magnetisation
for different bath temperatures, parameterised by the values of $\bar{y}_2$;
increasing values imply decreasing bath temperature. The other paramters
are set at $h_1=0, h_2=0.5, J_1=4.0, J_2=3.5, J_{12}=0.05$.}
\end{figure}

\end{document}